\documentclass[preprint2,tighten]{aastex6}
\usepackage{amsmath,amstext}
\usepackage[T1]{fontenc}
\usepackage{apjfonts}
\usepackage[figure,figure*]{hypcap}
\usepackage{float}

\newcommand{\be}{\begin{displaymath}}
\newcommand{\ee}{\end{displaymath}}
\newcommand{\bea}{\begin{eqnarray*}}
\newcommand{\eea}{\end{eqnarray*}}

\shorttitle{Obliquities of Kepler stars}
\shortauthors{Winn et al.}

\begin{document}

%
\def\ltsima{$\; \buildrel < \over \sim \;$}
\def\lsim{\lower.5ex\hbox{\ltsima}}
\def\gtsima{$\; \buildrel > \over \sim \;$}
\def\gsim{\lower.5ex\hbox{\gtsima}}
%

\title{Constraints on the Obliquities of {\it Kepler} Planet-Hosting Stars}

\author{Joshua N.\ Winn\altaffilmark{1}}
\author{Erik A.\ Petigura\altaffilmark{2,8}}
\author{Timothy D.\ Morton\altaffilmark{1}}
\author{Lauren M.\ Weiss\altaffilmark{3,9}}
\author{Fei Dai\altaffilmark{4,1}}
\author{Kevin C.\ Schlaufman\altaffilmark{5}}
\author{Andrew W.\ Howard\altaffilmark{2}}
\author{Howard Isaacson\altaffilmark{6}}
\author{Geoffrey W.\ Marcy\altaffilmark{6,10}}
\author{Anders Bo Justesen\altaffilmark{7}}
\author{Simon Albrecht\altaffilmark{7}}

\altaffiltext{1}{Princeton University, Princeton, NJ 08540, USA}

\altaffiltext{2}{California Institute of Technology, Pasadena, CA
  91125, USA}

\altaffiltext{3}{Institut de Recherche sur les Exoplan\`{e}tes, and
  Universit\'{e} de Montr\'{e}al, Montr\'{e}al, Canada}

\altaffiltext{4}{Massachusetts Institute of Technology, Cambridge, MA
  02139, USA}

\altaffiltext{5}{Johns Hopkins University, Baltimore, MD 21218, USA}

\altaffiltext{6}{University of California at Berkeley, Berkeley, CA
  94720, USA}

\altaffiltext{7}{Stellar Astrophysics Centre, Department of Physics
  and Astronomy, Aarhus University, Ny Munkegade 120, DK-8000 Aarhus
  C, Denmark}

\altaffiltext{8}{Hubble Fellow}

\altaffiltext{9}{Trottier Fellow}

\altaffiltext{10}{Professor Emeritus}

\begin{abstract}

  Stars with hot Jupiters have obliquities ranging from $0^\circ$ to
  $180^\circ$, but relatively little is known about the obliquities of
  stars with smaller planets. Using data from the California-{\it
    Kepler} Survey, we investigate the obliquities of stars with
  planets spanning a wide range of sizes, most of which are smaller
  than Neptune.  First, we identify 156 planet hosts for which
  measurements of the projected rotation velocity ($v\sin i$) and
  rotation period are both available.  By combining estimates of $v$
  and $v\sin i$, we find nearly all the stars to be compatible with
  high inclination, and hence, low obliquity ($\lesssim$20$^\circ$).
  Second, we focus on a sample of 159 hot stars ($T_{\rm
    eff}>$~6000\,K) for which $v\sin i$ is available but not
  necessarily the rotation period. We find six stars for which $v\sin
  i$ is anomalously low, an indicator of high obliquity.  Half of
  these have hot Jupiters, even though only 3\% of the stars that were
  searched have hot Jupiters. We also compare the $v\sin i$
  distribution of the hot stars with planets to that of 83 control
  stars selected without prior knowledge of planets. The mean $v\sin
  i$ of the control stars is lower than that of the planet hosts by a
  factor of approximately $\pi/4$, as one would expect if the planet
  hosts have low obliquities.  All these findings suggest that the
  {\it Kepler} planet-hosting stars generally have low obliquities,
  with the exception of hot stars with hot Jupiters.
    
\end{abstract}


\section{Introduction}
\label{sec:introduction}

One might expect good alignment between the rotation of a star and the
revolutions of its planets, and indeed, the Sun has a low obliquity.
Nevertheless, some exoplanetary systems have spin-orbit misalignments,
for reasons that remain unknown [see, e.g.,
  \citet{Triaud+2010,Albrecht+2012}, or the review by
  \citet{WinnFabrycky2015}]. Among the proposed reasons are a
primordial tilt of the protoplanetary disk \citep{Batygin+2012},
gravitational interactions between planets \citep{Chatterjee+2008},
Kozai-Lidov oscillations of a planetary orbit
\citep{FabryckyTremaine2007}, spin-orbit interactions between the star
and short-period planets \citep{SpaldingBatygin2014}, and angular
momentum redistribution within the star \citep{Rogers+2012}.  In
short, our ignorance is such that we do not know whether to blame the
disk, the planets, the star, or a neighboring star.

Most of our knowledge of obliquities is limited to stars with close-in
giant planets, with sizes $\gsim$6~$R_\oplus$ and orbital periods
$\lsim$10~days.  This is for practical reasons. Several of the
techniques for determining obliquities rely on transit signals, which
are easier to detect for close-in giant planets. Relatively little is
known about stars with smaller planets or wider-orbiting planets.
This is a major gap in our understanding because the {\it Kepler}
mission has revealed that close-in giant planets are rare in
comparison to systems of smaller and wider-orbiting planets
\citep{Howard+2012}. The {\it Kepler} sample of planetary systems is
dominated by planets smaller than 4~$R_\oplus$ with periods ranging
from 3 to 100~days.

Stellar obliquities have been measured in only a few of the {\it
  Kepler} planetary systems
\citep{Hirano+2012b,SanchisOjeda+2012,SanchisOjeda+2013,Chaplin+2013,Huber+2013,Campante+2016}.
Based on an analysis of five {\it Kepler} stars with multiple
transiting planets, all of which were found to have low obliquities,
\citet{Albrecht+2013} suggested that the high obliquities are confined
to hot-Jupiter hosts. Soon afterward, \citet{Huber+2013} found a high
obliquity for Kepler-56 \citep{Huber+2013}, which remains the only
system known to have two or more coplanar planets and a misaligned star.  In
that case, the misalignment may have been caused by the torque from a
wider-orbiting third planet \citep{Otor+2016,GratiaFabrycky2017}.

\citet{Mazeh+2015} made an important advance by measuring the
amplitude of photometric variability associated with rotation for a
large sample of {\it Kepler} stars.  Among the stars with effective
temperatures $\lsim$\,6000\,K, those without detected transiting planets
displayed a lower level of variability than stars with detected
planets.  The ratio was approximately $\pi/4$, as one would expect if
the planet-hosting stars have low obliquities and the others are
randomly oriented.  For hotter stars (6000-6500\,K) the variability
enhancement of transit hosts was not seen, suggesting that hot stars
have more random obliquities.  These results seemed to harmonize with
previous studies of hot-Jupiter hosts, which showed that hot stars
have a broader obliquity distribution than cool stars
\citep{Schlaufman2010,Winn+2010,Albrecht+2012}.  Thus, the variability
study suggested that hot stars have a broad range of obliquities
regardless of the properties of their planets, a potentially important
clue to the origin of spin-orbit misalignments.  The boundary of
$\approx$6000\,K\footnote{The temperature separating the different
  obliquity regimes was observed to be about 6250\,K by
  \citet{Winn+2010} and found to be $6090_{-110}^{+150}$\,K in a
  statistical analysis by \citet{Dawson2014}.} may be related to the
``Kraft break'' that distinguishes cool stars with thick convective
envelopes from hot stars with radiative envelopes \citep{Struve1930,
  Schatzman1962, Kraft1967}.

In this paper we report on further explorations of the obliquities of
{\it Kepler} stars, enabled by the California-{\it Kepler} Survey
[CKS: \citet{Petigura+2017,Johnson+2017}].  The CKS team performed
high-resolution optical spectroscopy of about a thousand stars with
transiting planets, provided a homogeneous catalog of spectroscopic
parameters, and clarified the masses and sizes of the stars and their
planets.  Of greatest importance for our work are the measurements of
the projected rotation velocity, $v\sin i$, for which the CKS team
demonstrated an accuracy of $\approx$1~km~s$^{-1}$
\citep{Petigura2015}.  We used these data in three different ways:
\begin{enumerate}

\item For stars with reliable determinations of $v\sin i$, stellar
  rotation period $P_{\rm rot}$, and stellar radius $R_\star$, it is
  possible to derive a constraint on the stellar inclination by
  comparing $v\sin i$ and $v\equiv 2\pi R_\star/P_{\rm rot}$.  This
  method has been employed by \citet{WalkowiczBasri2013},
  \citet{Hirano+2012}, \citet{Hirano+2014} and \citet{MortonWinn2014},
  among others. The latter authors found 2$\sigma$ evidence that stars
  with multiple transiting planets (``multis'') have lower obliquities
  than stars with only one detected transiting planets
  (``singles''). The CKS provides a larger, more accurate and more
  homogeneous dataset than was previously available.

\item High-obliquity systems can sometimes be recognized by virtue of
  an anomalously low $v\sin i$ for a star of a given type.  This is
  because the orbital inclination $i_{\rm o}$ must be near $90^\circ$
  for transits to occur, and a low obliquity for a transiting-planet
  host implies $\sin i\approx \sin i_{\rm o} \approx
  1$. \citet{Schlaufman2010} devised a statistic to quantify the
  meaning of ``anomalously low'' and computed it for a large sample of
  hot-Jupiter systems.  With the CKS data, we can apply this and
  related techniques to a larger and more diverse sample of planetary
  systems.

\item We can also test for systematically low obliquities by comparing
  the $v\sin i$ distribution of transiting-planet hosts with that of a
  sample of randomly oriented stars. Low-obliquity planet hosts must
  have $\sin i\approx 1$, whereas randomly oriented stars have
  $\langle \sin i \rangle = \pi/4$.  Thus, if the planet hosts have
  low obliquities, the mean $v\sin i$ of the randomly oriented stars
  should be lower than that of the planet hosts by a factor of
  $\pi/4$.

\end{enumerate}

We used these techniques to search for individual systems with high
obliquities, and to perform statistical comparisons between different
populations of planet-hosting stars.  We compared multis and singles,
stars with different types of planets, and hot vs.\ cool stars.  We
also tried to test the notion that hot stars have nearly random
obliquities regardless of the properties of their planets, as
suggested by the prior study of variability amplitudes.

These methods have some limitations that are common to any $\sin
i$-based technique.  First, they are unable to distinguish between
prograde and retrograde motion.  Second, because of the flattening of
the sine function near $90^\circ$, it is difficult to distinguish
between inclinations in the range $45^\circ$--$90^\circ$. Third, even
if the inclination is found to be near 90$^\circ$, the stellar
obliquity is not necessarily small.  This is because the inclination
is only one aspect of the obliquity $\theta$:
\begin{equation}
\cos\theta = \sin i~\cos\lambda~\sin i_{\rm o} +
\cos i~\cos i_{\rm o}
\approx \sin i~\cos\lambda,
\end{equation}
where $\lambda$ is the
sky-projected angle between the rotational and orbital axes.
Another useful relation between $\theta$ and $i$ is
\begin{equation}
\label{eqn:sini}
\cos i = \sin\theta \cos\phi,
\end{equation}
where $\phi$ is the azimuthal angle of the line of sight projected
onto the star's equatorial plane.
It is safe to assume that $\phi$ is uniformly distributed
in a sample of unrelated stars. This is what
gives $\sin i$-based techniques
the power to infer the obliquity distribution of a population of stars.

This paper is organized as follows. Section~\ref{sec:data} presents
the data and the properties of the stars and planets in our sample.
Section~\ref{sec:vvsini} considers the stars with measured rotation
periods, and compares estimates of the true and projected rotation
velocities.  Section~\ref{sec:lowvsini} presents two tests for
anomalously low values of $v\sin i$.  Section~\ref{sec:distributions}
compares the $v\sin i$ distributions of different samples of
planet-hosting stars with one another and with a sample of stars
selected without prior knowledge of any planets.
Section~\ref{sec:discussion} summarizes the results and their
relationship to the questions raised in this introduction, and
makes suggestions for future work.

\section{Data}
\label{sec:data}

\begin{figure*}[h!]
 \begin{center}
 \leavevmode
 \epsscale{0.8}
 \plotone{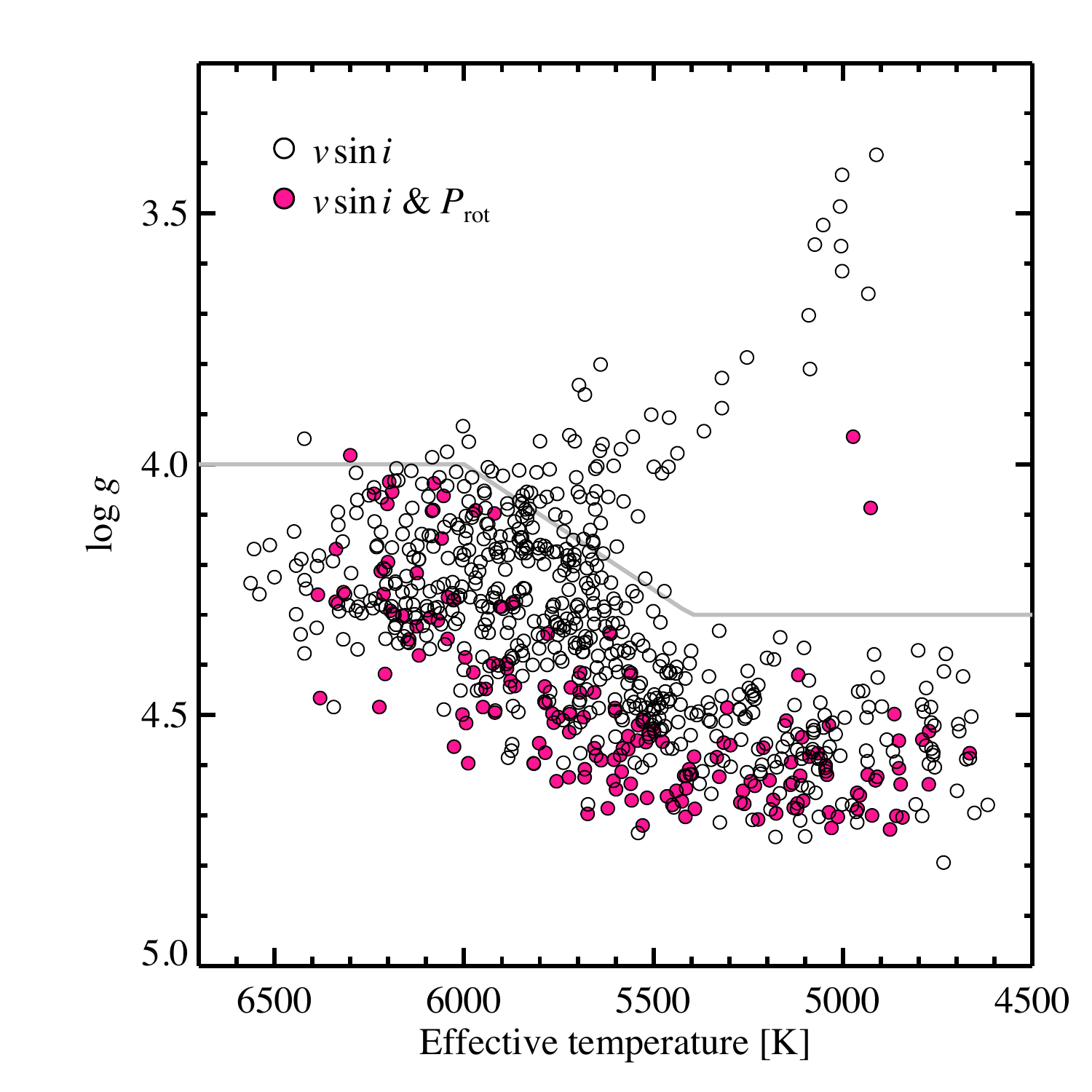}
 \end{center}
 \vspace{-0.15in}
 \caption{ The stars. Spectroscopic parameters are from
   \citet{Petigura+2017}. The colored points are those for which
   a reliable measurement of the rotation period is available.
   The stars below the gray line are those we consider ``dwarfs.''
  \label{fig:hr}}
\end{figure*}

\begin{figure*}[h!]
 \begin{center}
 \epsscale{0.9}
 \plotone{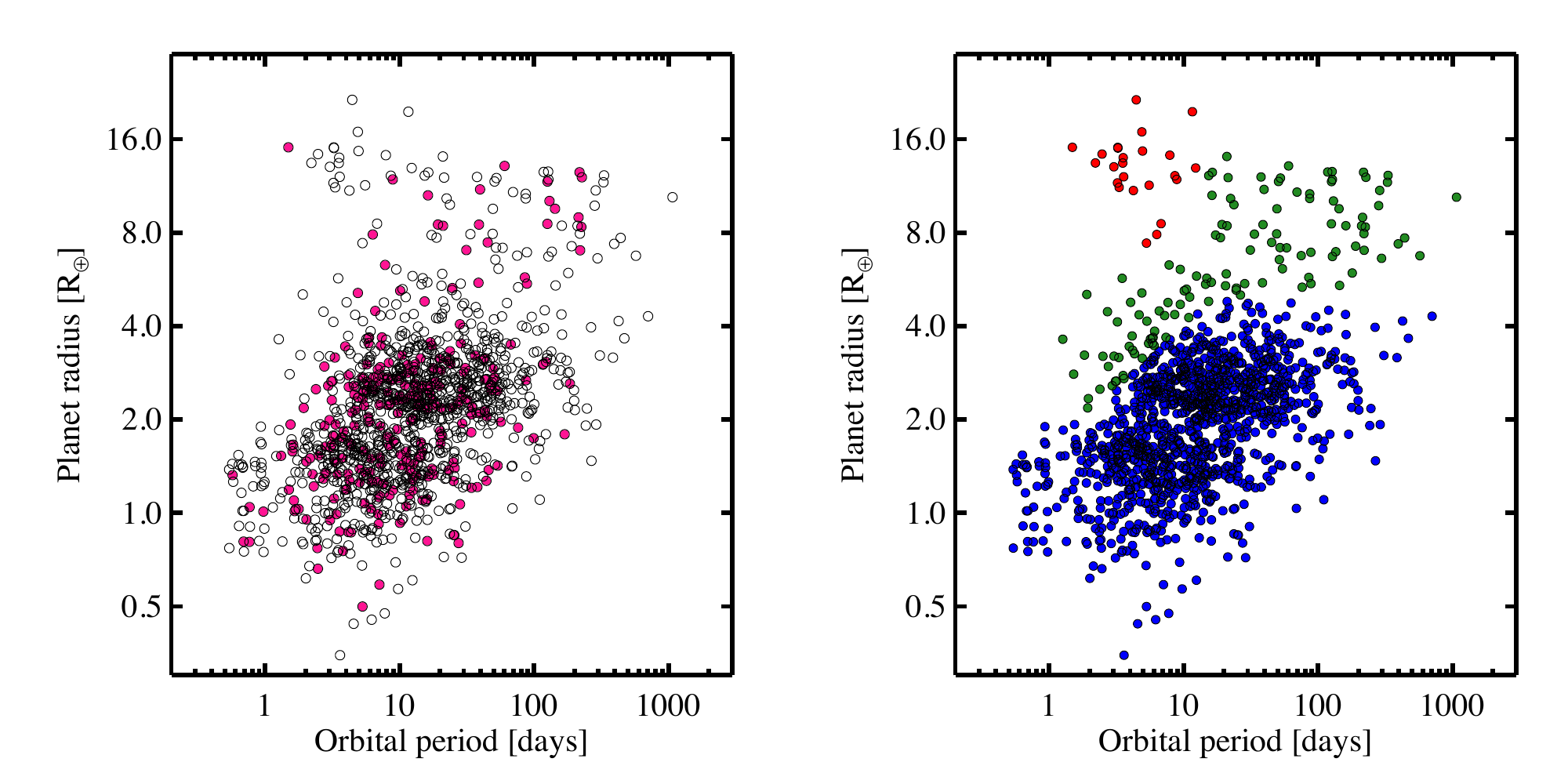}
 \end{center}
 \vspace{-0.1in}
 \caption{ The planets. {\it Left.}---Colored points are those for
   which a reliable measurement of the rotation period is available,
   as in Figure~\ref{fig:hr}. {\it Right.}---Same, but color-coded
   according to our chosen planet categories (see the
   text).  \label{fig:pr}}
\end{figure*}

\cite{Petigura+2017} presented spectroscopic parameters for 1305 stars
designated as {\it Kepler} Objects of Interest (KOIs).  These stars
have two things in common: they were selected as targets for the {\it
  Kepler} mission according to the criteria of \citet{Batalha+2010},
and at least one photometric signal was detected that resembles the
expected signal of a transiting planet. The spectroscopic parameters
were derived using {\tt SpecMatch} \citep{Petigura2015}.  This code
fits selected regions of an observed spectrum with a synthetic
spectrum.  The synthetic spectrum is generated by interpolating
between theoretical spectra in the library of \citet{Coelho+2005} and
convolving with broadening kernels for rotation, macroturbulence, and
instrumental resolution.  Macroturbulence is assumed to depend on
effective temperature, according to Equation~(1) of
\citet{ValentiFischer2005}.

For our study, we selected the 768 stars with planets designated as
``confirmed'' by \cite{Petigura+2017}, who based their assessment on
work by \citet{Morton+2016} and \cite{Mullally+2016}.  We omitted three
stars (KOI~935, 1060, and 1102) for which the $v\sin i$ measurement
was found to be unreliable, after testing for internal consistency
between the fits to different regions of the spectrum.  This left a
sample of 765 stars.

\citet{Petigura2015} gauged the accuracy of the CKS
projected rotation velocities through comparisons to measurements
based on the Rossiter-McLaughlin effect \citep{Albrecht+2012}.
For stars with $v\sin i>2$~km~s$^{-1}$, he found the accuracy to be
1~km~s$^{-1}$ or better.  For lower projected rotation
velocities he found that the results provide only an upper limit.
As a convenient interpolation between
these cases, we adopted uncertainties (in km~s$^{-1}$) of
\begin{equation}
1 + \frac{1}{(v\sin i/2)^4 + 1}.
\end{equation}

To assign rotation periods to CKS stars we consulted the catalogs of
\citet{Mazeh+2015} and \citet{Angus+2017}.\footnote{The published work
  of \citet{Angus+2017} does not include a table of rotation periods.
  T.~Morton furnished a list of periods from that study that are deemed
  reliable based on inject-and-recover simulations.} These groups used
two different techniques to detect quasiperiodic photometric
variations associated with the rotation of starspots or other
inhomogeneities on the stellar surface.  We considered a period to be
reliable if it appears in both catalogs with the same value to within
10\%.  There are 232 stars in our sample with a photometric period in
the catalog of \citet{Mazeh+2015}, of which 179 are also in the
catalog of \citet{Angus+2017} with a matching period.  From this
sample we omitted 23 stars for which \citet{Furlan+2017} found the
{\it Kepler} photometric aperture to contain a stellar companion with
a brightness within 3~mag of that of the intended target star.  In
such cases, it is not clear which star is producing the photometric
variations. This left a sample of 156 stars with reliably measured
photometric periods.  We examined all the light curves and confirmed
visually that the tabulated periods are reasonable.

Figure~\ref{fig:hr} shows the effective temperature and surface
gravity of all 765 stars, and identifies the 156 stars (20\% of the
total) with reliably measured photometric periods.  The stars with
periods have systematically higher surface gravity, an indication of
smaller size and younger age.  This is consistent with the well-known
tendency of young stars to be more active and spotted than older
stars.  The stars below the gray line are those we consider ``dwarfs''
in the sections to follow.  It is defined by two horizontal lines at
$\log g =4.0$ and 4.3, joined by a straight line between $T_{\rm
  eff}=$~5400\,K and 6000\,K.

Figure~\ref{fig:pr} shows the orbital period and radius of the largest
transiting planet belonging to each star.  The left panel identifies
the stars with reliable photometric periods.  The right panel assigns
colors to the data points based on categories that seem
astrophysically distinct and that we chose to track separately
throughout this study.  Red is for hot Jupiters, defined by the
criteria $R_{\rm p} > 7~R_\oplus$ and $P_{\rm orb} < 13$~days.  Green
is for wider-orbiting giant planets, as well as planets within the
``hot Neptune desert'' identified by \citet{Mazeh+2016}.  To qualify
for this category, the planet radius must either exceed $5~R_\oplus$
or the radius defined by the line connecting the points $(2.5,3)$ and
$(4,10)$ in radius-period space.  Blue is for the remaining planets,
constituting the bulk of the sample.

\section{Projected and true rotation velocities}
\label{sec:vvsini}

\begin{figure*}[h!]
 \begin{center}
 \leavevmode
 \epsscale{1.0}
 \plotone{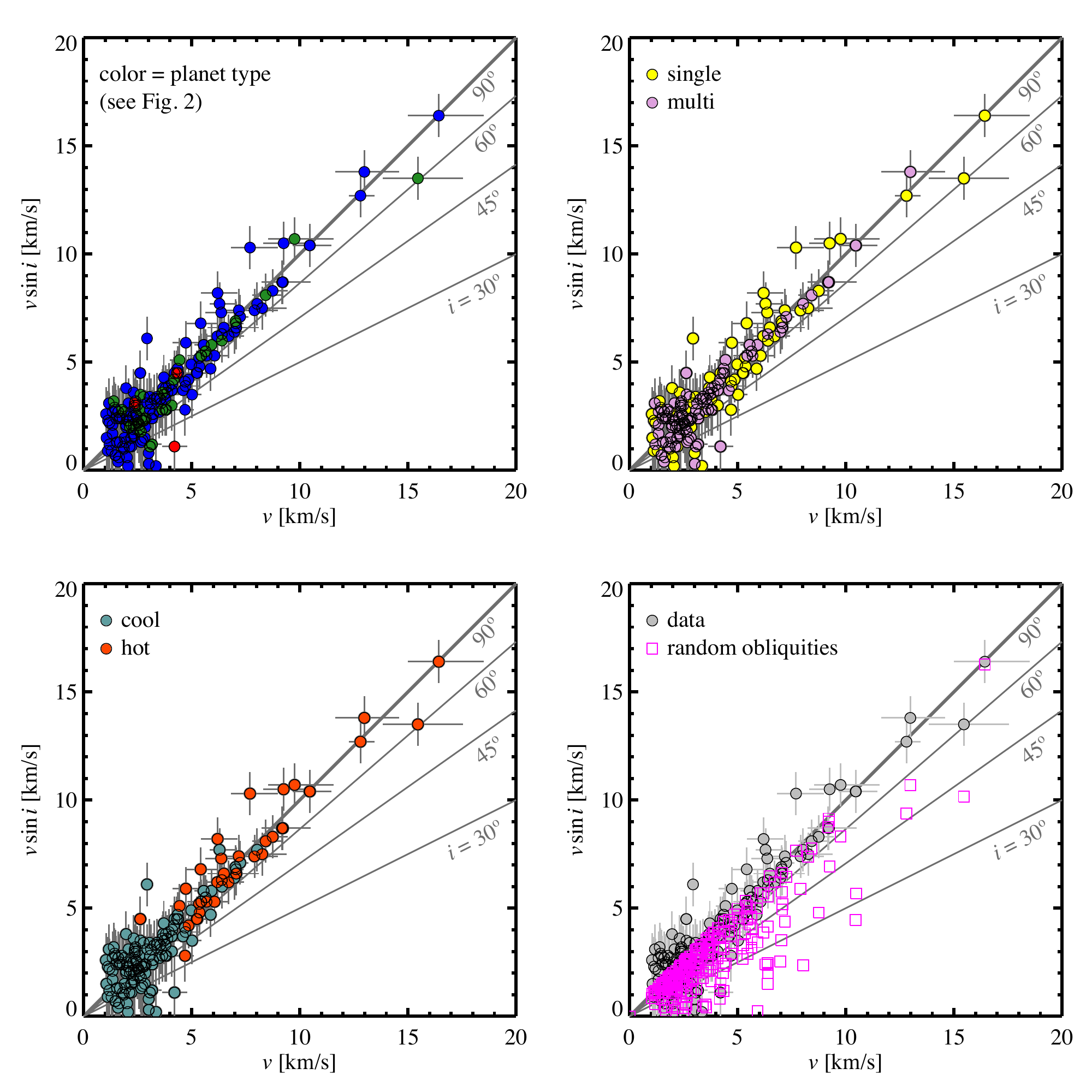}
 \end{center}
 \vspace{-0.25in}
 \caption{ Measured sky-projected rotation velocity versus calculated
   true rotation velocity. Any points falling below the identity line
   are candidate misaligned stars ($\sin i < 1$). {\it Upper
     left.}---Color indicates planet type, using the same scheme as in
   Fig.~\ref{fig:pr}.  {\it Upper right.}---Color indicates whether
   the star has more than one detected transiting planet.  {\it Lower
     left.}---Color indicates whether the star is hotter or cooler
   than 6000\,K.  {\it Lower right.}---Gray points are real
   data. The magenta squares are synthetic data with the same values of
   $v$ as real data and values of $\sin i$ chosen from a distribution of random
   orientations.
  \label{fig:vvsini_panels}}
\end{figure*}

For each star with a reliably measured photometric period, we assumed
that the photometric period is the stellar rotation period and
computed $v\equiv 2\pi R_\star/P_{\rm rot}$.  By using the same letter
$v$ that appears in $v\sin i$, we implicitly assumed they refer to the
same rotation velocity, i.e., we neglected systematic errors due to
differential rotation on both the measurement of $v\sin i$ from the
spectral lines and on the determination of $P_{\rm rot}$ from the
photometric variations.  These are expected to be 5-10\% effects
\citep{Hirano+2014} and to at least partly cancel out; differential
rotation causes both the inferred $v$ and $v\sin i$ to be biased
toward lower values than $v_{\rm eq}$ and $v_{\rm eq}\sin i$, where
$v_{\rm eq}$ is the equatorial rotation velocity.

Figure~\ref{fig:vvsini_panels} shows the results.  The different
panels use color to distinguish between different subsamples: by
planet type (upper left), single vs.\ multiple transiting planets
(upper right), and hot vs.\ cool stars (lower left).  The lower right
panel helps to put these results into perspective by showing synthetic
$v\sin i$ data for an isotropically oriented population of stars.  The
synthetic data were generated by adopting the values of $v$ from the
data, drawing $\cos i$ from a uniform distribution, and multiplying
$v$ by $\sin i$.

For velocities below about 4~km~s$^{-1}$, the large fractional
uncertainties make it is difficult to make any useful comparisons.
For higher velocities, the stars with reliable periods cluster around
the identity line. The standard deviation of $v\sin i - v$ is
1.0~km~s$^{-1}$, similar to the measurement uncertainty.  This implies
$\sin i \approx 1$.  Since the orbital inclination also has $\sin i_{\rm o} \approx 1$,
these results are consistent with (but do not
require) a low obliquity.

We used these data to establish lower limits on the inclination of
each star, focusing attention on the stars with $v > 4$~km~s$^{-1}$
for which meaningful constraints are possible.  We chose to express
the constraints as upper limits on $\cos i$. This facilitates the
interpretation because $\cos i$ is uniformly distributed for a
population of randomly oriented stars.
Following
\citet{MortonWinn2014}\footnote{We note that Equation~(13) of
  \citet{MortonWinn2014} has an error: it is missing the factor of $u$
  in the integral.}, the likelihood function is
\begin{equation}
\mathcal{L}({\rm data}\,|\,\cos i) = \int_0^\infty u p_1(u) p_2\left( \frac{u}{\sqrt{1 -
    \cos^2 i}} \right) du,
\end{equation}
where $p_1(u)$ and $p_2(u)$ are the likelihoods for $v$ and $v\sin i$,
respectively, based on the data. We assume $p_1$ and $p_2$ to be
Gaussian functions with means and standard deviations set by the
measured value and 1$\sigma$ uncertainties.

Table~\ref{tbl:stars} gives the results for each star.  The 95\%
confidence upper limits on $\cos i$ range from 0.5 to 0.8 for most stars,
with a mean of 0.7.  Thus, one way to summarize this investigation is
an unsuccessful search for any inclinations lower than about
$\cos^{-1}(0.7)$ or $45^\circ$.  These constraints are relatively
weak, for reasons explained in the introduction.  Even if a completely
random orientation is chosen for a given star, there is a 70\% chance
that $\cos i$ will be smaller than 0.7.

We may draw stronger conclusions about the ensemble of
stars. Visual inspection of the lower right panel of
Figure~\ref{fig:vvsini_panels} indicates that the data are incompatible
with an isotropic distribution of obliquities: there are not enough
low values of $v\sin i$.  To quantify this impression, we
compute the probability of drawing $N$ stars from an isotropic
distribution, each of which is observed to have $\cos i<z_i$:
\begin{equation}
\label{eqn:piso}
p_{\rm iso} = \prod_{i=1}^N z_i.
\end{equation}
For the 54 stars with $v>4$~km~s$^{-1}$, 
we find $p_{\rm iso} = 1.5\times 10^{-9}$.  For the subsample of 32 hot stars,
a sample we will discuss further in \S~\ref{sec:discussion}, we
find $p_{\rm iso} = 2\times 10^{-6}$.

We also constrained the obliquity distribution using a hierarchical
Bayesian method, as advocated by \citet{MortonWinn2014}.
We assumed the obliquities follow a Rayleigh distribution,
\begin{equation}
p(\theta) = \frac{\theta}{\sigma^2} \exp(-\theta^2/2 \sigma^2),
\end{equation}
and used a Monte Carlo Markov Chain to determine the posterior
distribution for $\sigma$, the mean obliquity.  For each proposed
value of $\sigma$, we drew $N$ obliquities from the corresponding
Rayleigh distribution, one for each star in the sample.  We also drew
$N$ azimuthal angles $\phi$ from a uniform distribution and calculated
$\sin i$ for each star using Eqn.~\ref{eqn:sini}.
The likelihood function was taken to be proportional to $\exp(-\chi^2/2)$ with
\begin{equation}
  \chi^2 = \sum_i^N \left[ \frac{v\sin i - v\times \sin i}
    {\sqrt{ \sigma_{v\sin i}^2 + \sigma_v^2 }} \right]^2.
\end{equation}
Here, the sum is over all the stars in the sample: $v$ and $v \sin i$
are the true and projected rotation rates, with uncertainties
$\sigma_{v\sin i}$ and $\sigma_v$; and $\sin i$ is the value assigned
through the Monte Carlo procedure described above.  We adopted a
uniform prior for $\sigma$.

For the sample of 58 stars with $v>4$~km~s$^{-1}$, we found
$\sigma<20^\circ$ with 99\% confidence.  Broader distributions do not
fit the data because they tend to produce too many low values of $\sin
i$.  Considering only the hot stars, cool stars, singles, and multis,
the upper limits are $22^\circ$, $36^\circ$, $24^\circ$ and
$26^\circ$, respectively. We find no evidence for any distinction
between these subsamples.

In interpreting these results we must remember that the stars are not
a random selection of ${\it Kepler}$ planet-hosting stars: they were
selected by virtue of having a robustly detectable photometric period.
Stars with low inclinations present lower-amplitude photometric
signals associated with rotation. This should cause a sample of stars
with detected signals to be deficient in low-inclination stars
compared to the broader sample of stars with transiting planets.
Suppose that the photometric variability amplitude varies as $\sin i$,
and that a reduction in the amplitude by a factor $f$ would have made
it impossible to detect the rotation period. Then the revised
probability that the stars belong to an isotropically oriented
population is
\begin{equation}
\label{eqn:piso-with-f}
p_{\rm iso} = \prod_{i=1}^N \frac{z_i}{\sqrt{1-f_i^2}}.
\end{equation}
The appropriate values of $f_i$ would need to be determined by modeling
the detection process for the rotation periods, which is beyond the scope
of this study.

Hence, we are not yet in a position to use these data to draw firm
conclusions about the obliquity distribution of planet-hosting stars
in general.  Despite this caveat, though, the conclusion that an
isotropic distribution is ruled out seems secure.  We find $p_{\rm
  iso} < 0.01$ even for $f=2/3$, i.e., the case in which the
photometric signals would have been undetectable had the amplitudes
been lower by only 33\%.

\section{Stars with anomalously low $v\sin i$}
\label{sec:lowvsini}

\begin{figure*}[ht!]
 \begin{center}
 \leavevmode \epsscale{0.9} \plotone{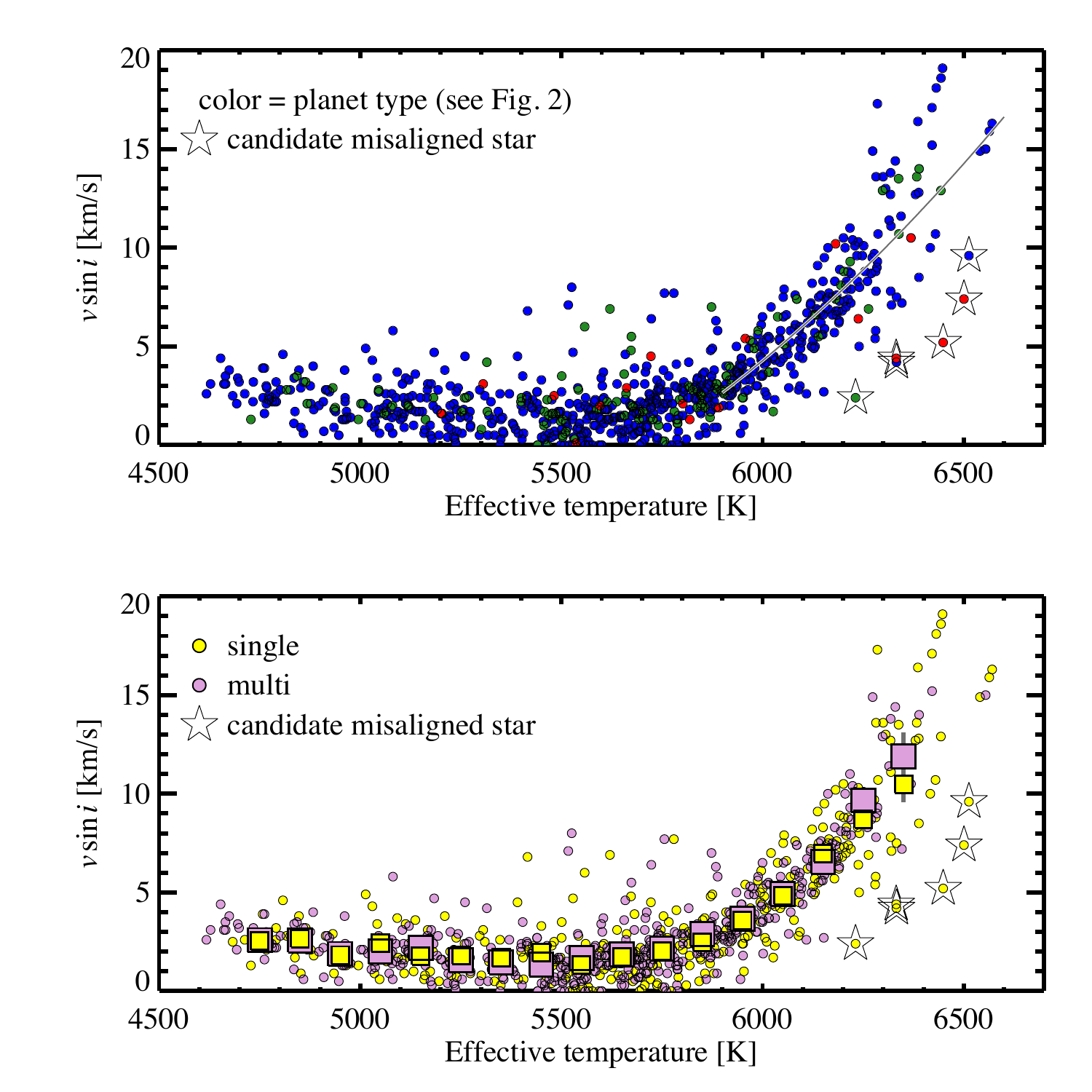}
 \end{center}
 \vspace{-0.25in}
 \caption{ Projected rotation velocities as a function of effective
     temperature. {\it Top.}---Color coded according to planet type,
   using the same categories as in Fig.~\ref{fig:pr}. The gray line is
   a fit to the data with $T_{\rm eff}>$~5900\,K. 
   The starred points are the most negative outliers from the
   fit, making them candidate misaligned stars.  {\it
     Bottom.}---Color coded to distinguish singles and multis. The
   large squares show the averages within 100\,K temperature bins.
  \label{fig:vt}}
\end{figure*}

Main-sequence stars of a given mass and age tend to have similar
rotation velocities.  This fact is the basis of gyrochronology, the
determination of a star's age from its observed rotation velocity
\citep{Barnes2003}.  It is also the basis of a method for identifying
stars being viewed at low inclination: such stars should have an
unusually low sky-projected rotation velocity. Figure~\ref{fig:vt}
shows the CKS measurements of $v\sin i$ as a function of effective
temperature, for all stars deemed ``dwarfs'' according to the boundary
line shown in Figure~\ref{fig:hr}. The points are color-coded to
convey the type of planet, and also whether the star has more than one
detected planet.

The rise in velocities with effective temperature, starting at around
6000\,K, is a manifestation of the Kraft break.\footnote{The slight
  decrease in velocity between 4600 and 5200\,K is harder to
  understand. To our knowledge this has not been observed before, and
  there is no evidence for such a trend in the large catalog of {\it
    Kepler} rotation periods \citep{McQuillan+2014}. Possibly, it is
  an artifact of adopting the simple relationship of
  \citet{ValentiFischer2005} between rotation and
  macroturbulence. These coolest stars play little to no role in our
  analyses.} Cooler stars have thick convective envelopes and develop
strong magnetic activity, allowing them to lose angular momentum
through magnetized winds.  Hotter stars lack this spin-down mechanism
and retain their initially rapid rotation rates. For stars cooler than
the Kraft break, the typical rotation velocity is only a few
km~s$^{-1}$, which is not much larger than the measurement
unceratinty.  This makes it impossible to identify cases of unusually
low $v\sin i$.  More useful for this technique are the stars hotter
than 6000\,K.  We searched this sample of 159 hot stars for
unusually low $v\sin i$ values in two ways.

\begin{figure*}[h!]
 \begin{center}
 \leavevmode
 \epsscale{0.75}
 \plotone{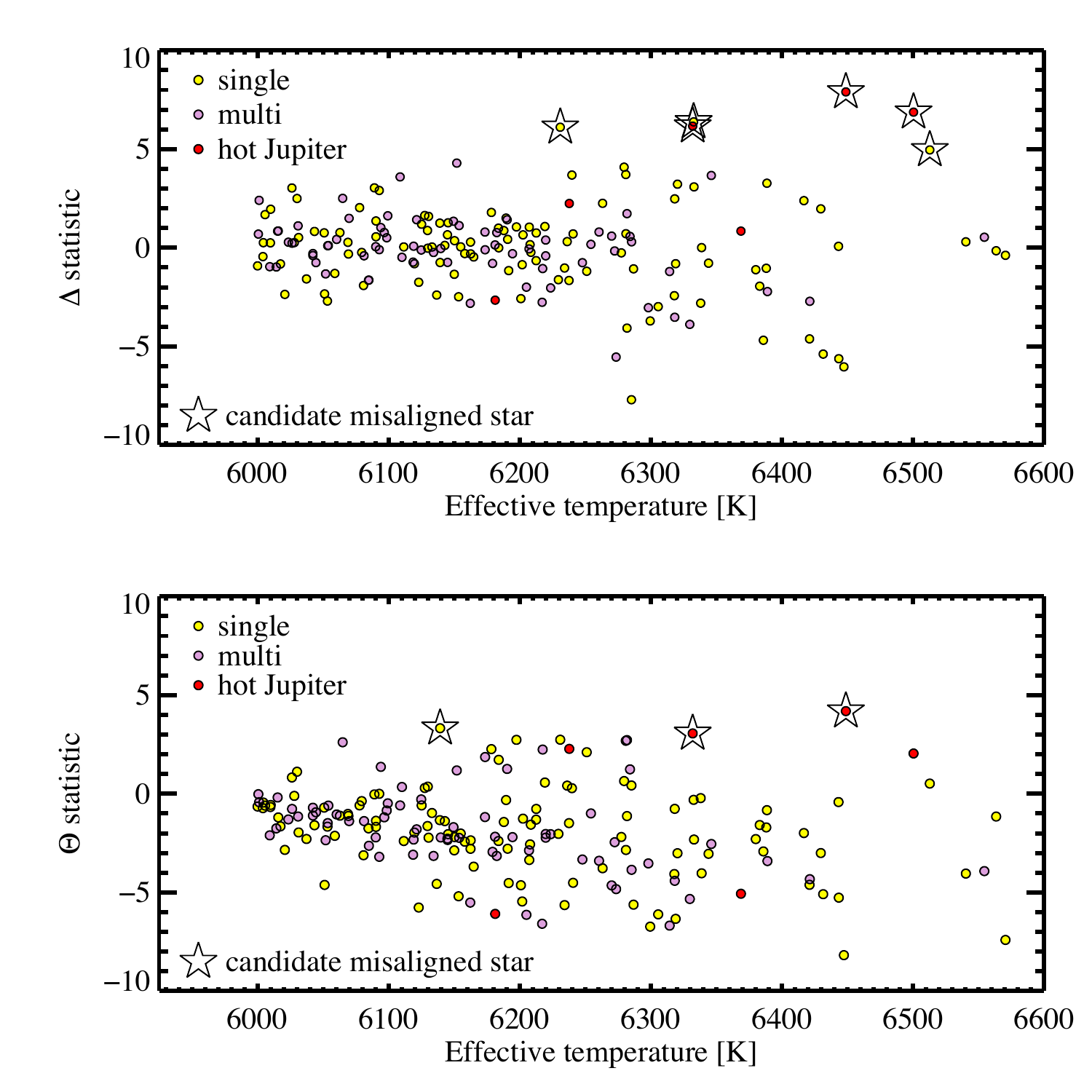}
 \end{center}
 \vspace{-0.25in}
 \caption{ Rotation statistics. {\it Top.}---Deviation $\Delta$
   from the best-fitting quadratic function of $v\sin i$ versus
   effective temperature. Large $\Delta$ implies an anomalously slow
   $v\sin i$. The starred points are the same as those in
   Fig.~\ref{fig:pr}. {\it Bottom.}---The Schlaufman~(2010) $\Theta$ statistic,
   which is high when the star appears to be rotating anomalously slowly.
   The starred points are those with $\Theta$ above the threshold of 2.9.
     \label{fig:deltatheta}}
\end{figure*}

The first method was to identify the largest outliers from the overall
trend of rising $v\sin i$ with effective temperature.  We fitted the
$v\sin i$ data with a quadratic function of $T_{\rm eff}$, and
computed the normalized residual,
\begin{equation}
  \Delta \equiv \frac{ (v\sin i)_{\rm fit} - (v\sin i)_{\rm obs} } {\sigma_v}
\end{equation}
for each star. Then we identified the systems with the largest values
of $\Delta$.  There are six outliers with $\Delta > 5$, identified by
the starred points in Figures~\ref{fig:vt} and \ref{fig:deltatheta}.
All six of these stars have only one detected transiting planet. Three
of them are hot-Jupiter hosts: KOI~2 (HAT-P-7 or Kepler-2), KOI~18
(Kepler-5), and KOI~98 (Kepler-14).  This is remarkable because there
are only five hot-Jupiter hosts in the entire sample.  If we were to
select six stars randomly from the sample of 159 hot stars, the chance
of selecting at least three hot-Jupiter hosts is $6\times 10^{-4}$.
This suggests that among the hot stars, those with hot Jupiters are
more likely to have high obliquities than those without hot Jupiters.

The obliquity of HAT-P-7 was already known to be large, thanks to
observations of the Rossiter-McLaughlin effect
\citep{Narita+2009,Winn+2009} and rotational splittings of
asteroseismic $p$-mode frequencies \citep{Benomar+2014,Lund+2014}.
HAT-P-7 and Kepler-5 were also flagged by \citet{Schlaufman2010} as
likely misaligned, based on their low values of $v\sin i$.  We are not
aware of any direct determinations of the obliquity of Kepler-14.

The other three stars with anomalously low $v\sin i$ are KOI~167
(Kepler-480), KOI~1117 (Kepler-774), and KOI~1852 (Kepler-982), which
host planets of size 2--3~$R_\oplus$ and orbital periods of 5--15 days.
These are good candidates for follow-up observations to test for a
high obliquity.
 
Our second method for identifying stars with anomalously low $v\sin i$
was to compute the rotation statistic $\Theta$ devised by
\citet{Schlaufman2010}, which compares the observed $v\sin i$ to the
expected rotation velocity $v$ for the given type of star. The inputs
were the star's $v\sin i$ from \citet{Petigura+2017}, and the stellar
mass and age estimated by \citet{Johnson+2017} by fitting
stellar-evolutionary models to the observed spectroscopic
parameters. The output, shown in Figure~\ref{fig:deltatheta}, is
roughly the number of sigma by which $v\sin i$ is lower than expected.
\citet{Schlaufman2010} calibrated the statistic by applying it to the
sample of \citet{ValentiFischer2005}, finding that stars with
$\Theta\gsim 2.9$ are likely to be misaligned.  By this criterion the
following systems are flagged as misaligned: KOI~2 (HAT-P-7), KOI~18
(Kepler-5), and KOI~2904 (Kepler-1382).

The first two members in this list were also flagged by the $\Delta$
statistic.  The last member, KOI~2904, has $v\sin i = 7.5$~km~s$^{-1}$
and $T_{\rm eff}=6139$~K.  It does not stand out in
Figure~\ref{fig:vt}.  It is nevertheless assigned a large $\Theta$ of
3.3 because the spin-down model of \citet{Schlaufman2010} predicts a
rotation period of 4.3 days, which, when combined with the stellar
radius of 1.94~$R_\odot$ leads to an expected $v=24$~km~s$^{-1}$. This
prediction is probably faulty, though. Although the star is classified as a
``dwarf'' according to the simple boundary drawn in
Figure~\ref{fig:hr}, the stellar-evolutionary modeling of
\cite{Johnson+2017} suggests it has begun evolving into a subgiant,
and has likely slowed its rotation as it has expanded. This type of
evolution is not taken into account in the spin-down model. Work is
underway by K.~Schlaufman on a revised statistic in which the model
for rotational evolution can accommodate somewhat evolved stars.

\section{Comparing $v\sin i$ distributions}
\label{sec:distributions}

We also sought evidence for differences in the $v\sin i$ distributions
between groups of stars. Any such differences might be attributed to
differences in obliquity.  An ideal basis for comparison would be a
large sample of $v\sin i$ measurements of stars spanning the same
range of spectroscopic parameters as the CKS stars that could be
safely assumed to be randomly oriented and to share the same
distribution of rotation velocities as the planet hosts.  One could
use the $v\sin i$ distribution of this control sample to determine the
intrinsic rotation velocity distribution of the relevant population of
stars \citep{ChandrasekharMunch1950}. This could then be compared to
the $v\sin i$ distribution of the hosts of transiting planets, to
learn about the distribution of $\sin i$.

An attractive possibility is the sample of stars analyzed by
\citet{ValentiFischer2005} and \citet{Brewer+2016}, in the program
entitled Spectral Properties of Cool Stars (SPOCS). The SPOCS stars
were chosen to be targets for a Doppler exoplanet survey, and have
been observed with the same instrument (Keck/HIRES) as the CKS stars.
In almost all cases the choice to perform high-resolution spectroscopy
was made without any prior knowledge of exoplanets, transiting or
otherwise.  Thus, it would seem that the spatial orientation of the
stars should be random, as desired.  A problem, though, is that the
surveyors excluded stars for which the available information indicated
possible problems with precise Doppler observations.  They avoided
close binaries.  They also rejected young and chromospherically active
stars based on published activity indicators ($S$ and $R'_{\rm HK}$),
X-ray fluxes, cluster-based ages, and lithium abundances.  The
resulting sample is therefore biased to some degree against rapid
rotators, calling into question the assumption that the control stars
and and planet-hosting stars have the same intrinsic distribution of
rotation velocities.

\begin{figure*}[ht!]
 \begin{center}
 \leavevmode
 \epsscale{0.9}
 \plotone{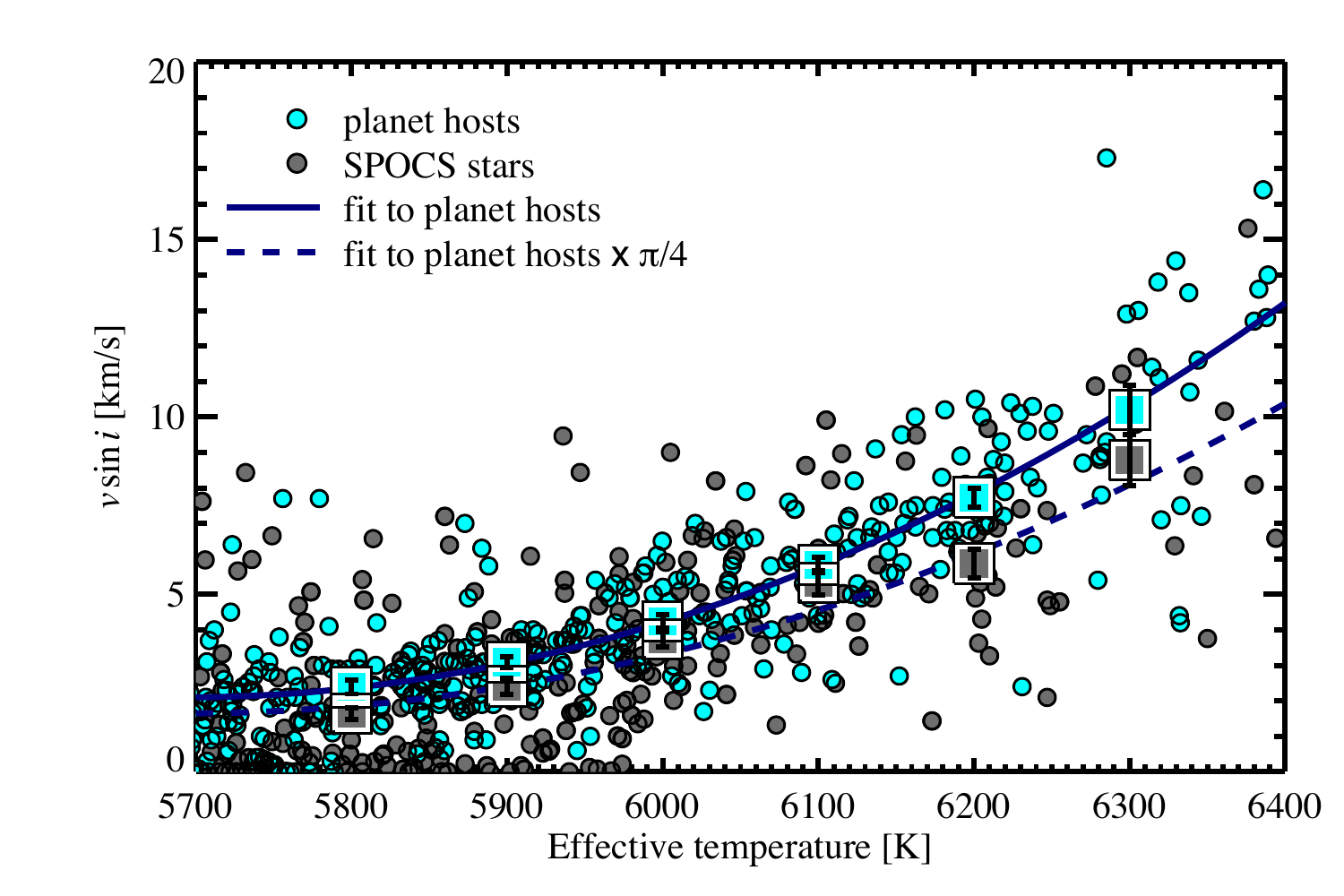}
 \end{center}
 \vspace{-0.25in}
 \caption{ Comparison with a control sample of stars from the SPOCS
   catalog, assumed to be randomly oriented. The squares are
   averages within 100\,K temperature bins.  At a given effective
   temperature, the mean $v\sin i$ of the control stars appears to be
   lower than that of the planet-hosting stars by a factor of
   approximately $\pi/4$.  This suggests that the planet hosts have low
   obliquites.
  \label{fig:compare_control}}
\end{figure*}

Despite this flaw, we decided to try this comparison because the SPOCS
stars come closer to the ideal than any other sample we were able to
find.  For consistency, we redetermined the $v\sin i$ values of the
SPOCS stars using the same version of the SpecMatch software that was
used on the CKS spectra. Figure~\ref{fig:compare_control}
compares the projected rotation velocities of the planet-hosting stars
(CKS) and control stars (SPOCS).  Shown are the data for individual
stars, as well as the averages within 100\,K temperature bins.  We
excluded the hot-Jupiter hosts from the averages because such stars
are already known to have a broad range of obliquities, and we wanted
to probe the obliquities of hot stars with other types of planets.
Assuming (i) the CKS and SPOCS stars have identical distributions of
rotation velocities, (ii) the CKS stars have low obliquities, and
(iii) the SPOCS stars are randomly oriented, we should observe that
the mean $v\sin i$ of the SPOCS stars is lower than that of the CKS
stars by a factor of $\pi/4$.

The data are compatible with these assumptions.  The solid line in
Figure~\ref{fig:compare_control} is a quadratic fit to the CKS data,
providing an estimate of the mean $v\sin i$ as a function of effective
temperature.  The dashed line is the same function after
multiplication by $\pi/4$, which gives a reasonable fit to the SPOCS
data.  In particular, the ratios of mean $v\sin i$ in the hottest two
temperature bins are $0.759\pm 0.059$ and $0.861\pm 0.091$, which are
both consistent with $\pi/4\approx 0.785$.  Thus, despite well-founded
concerns about the control sample, there is suggestive evidence that
hot stars lacking hot Jupiters have low obliquties.

Independently of the control sample, we searched for differences
between the $v\sin i$ distributions of different subsamples of the
planet-hosting stars.  Since it has been suggested that the multis
have systematically lower obliquities than the singles
\citep{MortonWinn2014}, we binned the $v\sin i$ data in temperature
for these two populations separately.  These binned results for the
singles and multis are shown with large squares in the bottom panel of
Figure~\ref{fig:vt}.  There are no statistically significant
differences, except for perhaps the two hottest temperature bins
($>$6200~K), within which the mean $v\sin i$ of the multis is higher
than that of the singles by 0.9 and 0.5$\sigma$.  Thus, this test
revealed no compelling difference between singles and multis.

We also tried comparing the distributions of the $\Delta$ and $\Theta$
statistics described in the previous section. Both of these statistics
are meant to quantify the difference between the observed $v\sin i$
and the rotation velocity one would expect for a star of the given
type.  Two-sided Kolmogorov-Smirnoff tests do not reveal any
significant differences between multis and singles ($p=0.4$ and 0.6
for $\Delta$ and $\Theta$, respectively). Likewise, there do not appear
to be any discernible differences between hosts of different planet
types, apart from hot Jupiters.
\newpage
\section{Summary and Discussion}
\label{sec:discussion}

\begin{figure}
 \begin{center}
 \leavevmode
 \epsscale{1.2}
 \plotone{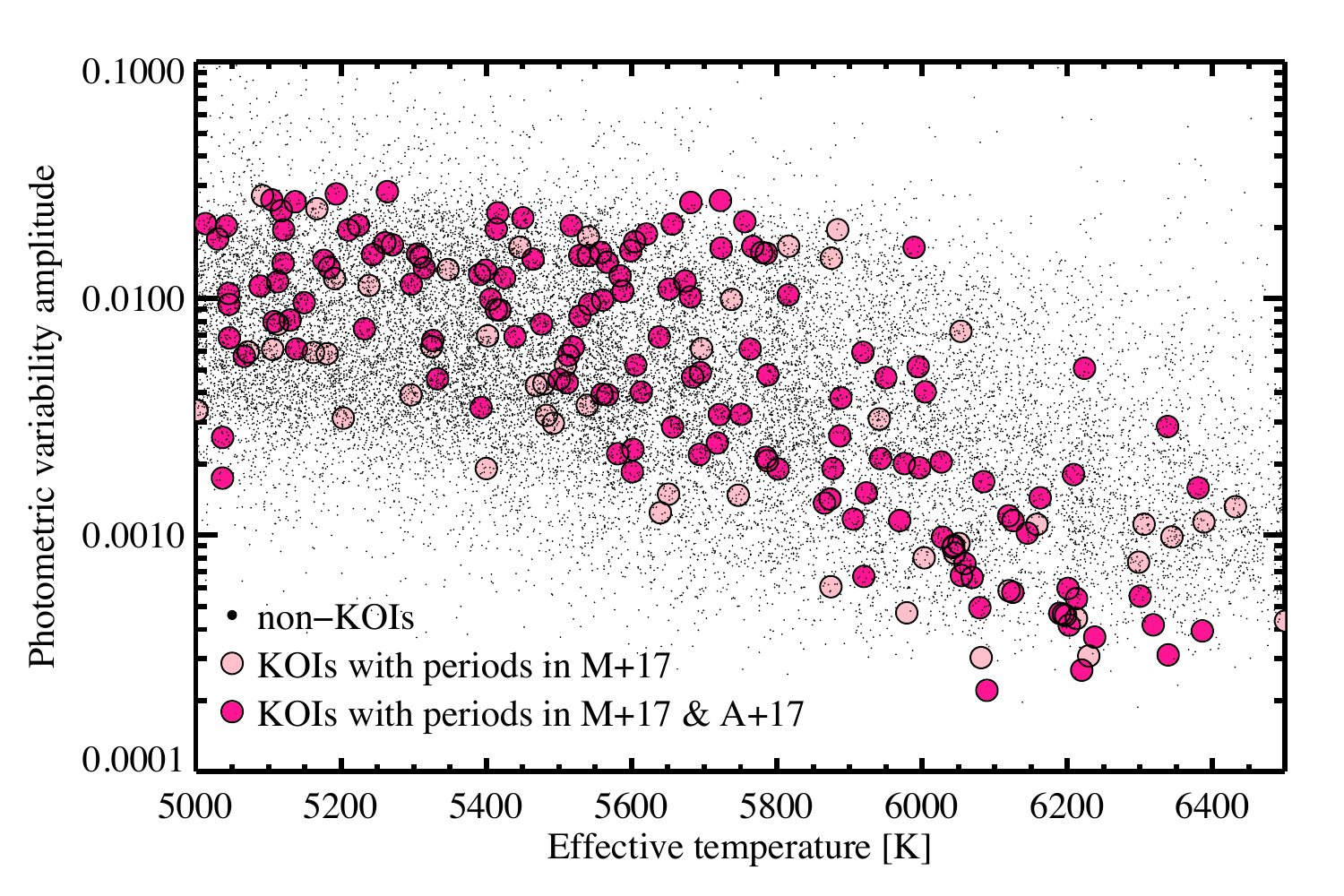}
 \end{center}
 \vspace{-0.25in}
 \caption{ Photometric variability amplitudes.  The small points are
   for stars without any detected transiting planets. The open circles are
   stars with transiting planets for which \citet{Mazeh+2015} detected
   the rotation period, and the solid red symbols are stars for which
   \citet{Angus+2017} agree on the period. This subset, analyzed in
   \S~\ref{sec:vvsini}, does not appear to be biased toward higher
   variability (higher $\sin i$) than the other planet-hosting stars
   in the same temperature range.
     \label{fig:rvar}}
\end{figure}

Using the newly available CKS data, we have investigated the obliquity
distribution of {\it Kepler} stars by comparing measurements of $v$
and $v\sin i$, seeking evidence for anomalously low $v\sin i$ values,
and testing for systematic differences between the $v\sin i$
distributions of different groups of stars.

Among the stars with reliably measured photometric periods, we found
no evidence for high obliquities.  When modeled as a Rayleigh
distribution, the mean obliquity is smaller than 20$^\circ$ with 99\%
confidence.  An isotropic distribution is strongly ruled out, both for
the entire sample, and for the subset of hot stars ($T_{\rm
  eff}>$~6000\,K). One reason this is interesting is that hot stars
with hot Jupiters are known from prior observations to have a very
broad obliquity distribution.  Not as much was known about hot stars
without hot Jupiters; our findings suggest they tend to have lower
obliquities.

\citet{WalkowiczBasri2013} and \citet{Hirano+2014} also sought
possible cases of spin-orbit misalignment through discrepancies
between $v\sin i$ and $v$, based on the data available at the time.
Table~\ref{tbl:prev-cands} summarizes our results for the objects that
they highlighted as possibly misaligned.  The column labeled
$N_\sigma$ is the number-of-sigma by which $v$ exceeds $v\sin i$,
which is smaller than 2 in all cases. Thus, we do not find compelling
evidence for misalignments in these systems based on the CKS data,
although it still may be worth follow-up observations to determine the
obliquities with other techniques.

We can also compare our results directly to the study of
\citet{Mazeh+2015}.  As mentioned in \S~\ref{sec:introduction} their
insight was to compare the photometric variability amplitudes of the
stars with transiting planets (KOIs) and the much larger sample of
stars without known transiting planets (non-KOIs).
Figure~\ref{fig:rvar} reproduces this comparison over the temperature
range 5000-6500\,K.  Among the cool stars, the KOIs show stronger
variability than the non-KOIs.  More puzzling is that for hot stars,
the KOIs are {\it less} variable than the non-KOIs.  An interpretation
purely in terms of obliquities is problematic.  If photometric
variability scales with $\sin i$, one would think that the obliquity
distribution producing the lowest level of variability would be an
isotropic distribution.  Even lower variability would require
preferential alignment of stellar rotation axes with the line of
sight, which seems implausible.  How, then, could the KOIs have a
lower level of variability than the non-KOIs, which presumably have an
isotropic distribution?

At least part of the explanation is a selection effect: the sample of
stars with detected transiting planets is biased toward lower
variability because it is more difficult to detect planets around
highly variable stars. The bias should be most pronounced for hot
stars because they are larger, causing the transit signals to be
smaller and closer to the detection threshold.  \citet{Mazeh+2015}
modeled the selection process and found that the level of bias could
account for about half of the difference in the observed variability
between KOIs and non-KOIs. This left open the possibility that the
variability of hot KOIs and non-KOIs is indeed nearly the same, which
would imply that hot stars have nearly random obliquities.  This would
be an important result because, as noted above, almost all the prior
work was restricted to stars with hot Jupiters.  If the high
obliquities are a more general phenomenon, this would point toward
theories involving mainly the star, rather than the disk, planetary
dynamics, or spin-orbit resonances.

However, our investigation of these stars (\S~\ref{sec:vvsini}) is not
compatible with this interpretation of the variability data.  The
sample contains 30 hot stars with reliably measured photometric
periods that are all compatible with high inclination and low
obliquity; they cannot be drawn from an isotropic distribution.  While
it is true that these 30 stars are only a subset of those examined by
\citet{Mazeh+2015}, there does not appear to be any reason for them to
be relatively biased toward high inclination.  Figure~\ref{fig:rvar}
shows that they have the same mean variability level as the larger
sample.

We placed an upper limit on the fraction of the hot stars in our sample
that could have been drawn from an isotropic distribution
in the following manner:
\begin{enumerate}

  \item Randomly select a subset $n$ of the hot stars.

  \item Calculate $p_{\rm iso}$ based only on these $n$ stars.

  \item Repeat the preceding steps 100 times to obtain the mean
    value of $p_{\rm iso}$ for that choice of $n$.

  \item Repeat the preceding steps for all $n$ ranging from 1 to 30,
    and find $n_{\rm max}$ for which $p_{\rm iso}<0.01$.

\end{enumerate}
The result is $n_{\rm max} = 11$ out of 33, i.e., fewer than one-third
of the hot stars are drawn from a randomly oriented distribution.
Given this result, we consider the interpretation of the
variability-amplitude data for hot stars to remain murky.

In the second part of the study, we identified some individual stars
that have unusually low projected rotation rates: KOI~2 (HAT-P-7),
KOI~18 (Kepler-5), KOI~98 (Kepler-14), KOI~167 (Kepler-480), KOI~1117
(Kepler-774), and KOI~1852 (Kepler-982). These are good candidates for
high-obliquity stars. Three of these are hot-Jupiter hosts and indeed
one of them (HAT-P-7) was already known to have a high obliquity.
These results support the notion that high obliquities are more common
among stars with hot Jupiters, compared to those hosting other types
of planets.  We also compared the $v\sin i$ distributions between
different samples of KOIs using several statistics, and found no
significant differences.

In neither part of our study did we find evidence that the multis have
lower obliquities than the singles, a trend that had been noted by
\citep{MortonWinn2014}.  Our non-confirmation of this result leads us
to suspect the result was spurious.  The trend was seen with only
2$\sigma$ confidence, and was based on a study of a smaller number of
stars (75) and a more heterogeneous dataset (drawn from 5 different
sources).

An important lesson we drew from this study is that hot stars lacking
hot Jupiters generally seem to have low obliquities.  This is the
converse of what was already known: hot stars with hot Jupiters tend
to have high obliquities \citep{Winn+2010,Albrecht+2012}.  Together
these findings suggest that the high obliquities are related to the
presence of close-in giant planets: the planet is somehow to blame for
the misalignment, or is at least associated with the causes of
misalignment. This argues against theories in which high obliquities
are a consequence of star-disk misalignment \citep{Batygin+2012} or of
processes taking place wholly within the star \citep{Rogers+2012}.
Instead, this result points toward a theory that requires a close-in
giant planet, while also making a distinction between hot stars and
cool stars.

In the theory of \citet{SpaldingBatygin2015}, spin-orbit misalignments
erupt from a resonance between the precession of the stellar spin axis
induced by a protoplanetary disk, and the precession of the disk
induced by a distant stellar companion.  Cool stars are able to
realign with their disks through magnetic torques, but hot stars
cannot because of their weaker and more disorderly magnetic fields.
In this scenario, hot stars should be misaligned with the orbits of
all the planets that ultimately form within the disk, and not just hot
Jupiters.  This does not seem compatible with our results. In addition
this mechanism cannot explain those few cool stars that are known to
have high obliquities, such as HD~80606 \citep{Winn+2009b,
  Hebrard+2010}, WASP-8 \citep{Queloz+2010}, and HAT-P-11
\citep{Winn+2010b,Hirano+2011}. One would need to invoke a separate
mechanism for such cases, such as Kozai-Lidov oscillations
\citep{FabryckyTremaine2007}.

\citet{MatsakosKonigl2015} proposed an interesting alternative: (i)
stars generally become misaligned with their disks (and hence the
planetary orbits) due to torques from nearby stars; (ii) many stars
also ingest a hot Jupiter early in their lifetimes, causing them to
realign with the planetary orbital plane; and (iii) this realignment
cannot be achieved for hot stars because of their higher mass and
angular momentum. This theory, as desired, distinguishes between hot
and cool stars and requires a close-in giant planet. However, in this
story the guilty planet no longer exists.  Our findings suggest that
the high obliquities are characterstic of stars with a currently
existing hot Jupiter.

Even after the infusion of new data from the California-{\it Kepler}
Survey, we have only limited information about the obliquities of the
{\it Kepler} planet-hosting stars.  As we emphasized in
\S~\ref{sec:vvsini}, our ability to draw general conclusions would be
enhanced by a reliable quantitative model for the selection function
associated with the detection of the photometric rotation period.  The
current sample is biased toward low obliquity to some degree, because
of the requirement that the photometric variations must be robustly
detected.

Another path forward would be to improve upon the existing control
sample of stars.  Ideally, we would measure the $v\sin i$ distribution
of a sample of at least several hundred {\it Kepler} stars selected
without regard to rotation velocity, orientation, or planet detection,
spanning the same range of spectroscopic parameters as the planet
hosts. The SPOCS sample that we employed in \S~\ref{sec:distributions}
was not designed for this purpose. It is smaller than the sample of
planet-hosting stars, and relatively deficient in hot stars.
Furthermore, at a given effective temperature it is probably biased
against rapid rotators due to a selection against young and
chromospherically active stars.

Finally, it is still valuable to perform Rossiter-McLaughlin
observations, analyze spot-crossing events, and perform other tests of
obliquities in individual systems. This remains difficult for the
relatively faint {\it Kepler} stars, but will be much easier with
stars that are 2-3~mag brighter, as we hope will be found by NASA's
forthcoming {\it Transiting Exoplanet Survey Satellite}
\citep{Ricker+2015}.

\acknowledgements We are grateful to the other CKS team members, and
the NASA {\it Kepler} team, for producing the database upon which this
study is based.
E.A.P.\ acknowledges support from Hubble Fellowship grant
HST-HF2-51365.001-A awarded by the Space Telescope Science Institute,
which is operated by the Association of Universities for Research in
Astronomy, Inc.\ for NASA under contract NAS\,5-26555.
S.A.\ and A.B.J.\ acknowledge support from the Danish
Council for Independent Research, through a DFF Sapere Aude Starting
Grant nr.\ 4181-00487B. 
The authors also wish to recognize and acknowledge the very
significant cultural role and reverence that the summit of Maunakea
has always had within the indigenous Hawaiian community. We are most
fortunate to have the opportunity to conduct observations from this
mountain.

\bibliographystyle{yahapj}
\bibliography{references}

\begin{deluxetable*}{ccccccccc}
\tabletypesize{\small}
\tablecaption{Stars with reliable rotation periods\label{tbl:stars}}
\tablewidth{0pt}

\tablehead{
  \colhead{KOI} &
  \colhead{KIC} &
  \colhead{$T_{\rm eff}$} &
  \colhead{$\log g$} &
  \colhead{Radius} &
  \colhead{Rotation Period} &
  \colhead{$v$} &
  \colhead{$v\sin i$} &
  \colhead{$\cos i_{\rm max}$} \\
  \colhead{number} &
  \colhead{number} &
  \colhead{[K]} &
  \colhead{[cgs]} &
  \colhead{[$R_\odot$]} &
  \colhead{[days]} &
  \colhead{[km~s$^{-1}$]} &
  \colhead{[km~s$^{-1}$]} &
  \colhead{(95\% conf.)}
}
\renewcommand{\arraystretch}{1.0}
\startdata
    49 &    9527334 &  $ 5779\pm 70$ & $   4.338\pm 0.10$ & $ 1.08^{+0.14}_{-0.09}$ & $   8.665\pm    0.075$ & $   6.421\pm     0.62$ & $    7.70\pm     1.00$ & $0.54$ \\
    85 &    5866724 &  $ 6219\pm 70$ & $   4.213\pm 0.10$ & $ 1.43^{+0.23}_{-0.17}$ & $   7.889\pm    0.210$ & $   9.313\pm     1.14$ & $    8.70\pm     1.00$ & $0.67$ \\
   107 &   11250587 &  $ 5919\pm 70$ & $   4.098\pm 0.10$ & $ 1.62^{+0.29}_{-0.22}$ & $  17.499\pm    0.800$ & $   4.747\pm     0.56$ & $    3.90\pm     1.06$ & $0.83$ \\
   203 &   10619192 &  $ 5722\pm 70$ & $   4.534\pm 0.10$ & $ 1.03^{+0.08}_{-0.05}$ & $  12.161\pm    0.111$ & $   4.367\pm     0.26$ & $    4.50\pm     1.04$ & $0.73$ \\
   257 &    5514383 &  $ 6162\pm 70$ & $   4.302\pm 0.10$ & $ 1.29^{+0.18}_{-0.13}$ & $   7.900\pm    0.081$ & $   8.400\pm     0.94$ & $    7.50\pm     1.01$ & $0.70$ \\
   271 &    9451706 &  $ 6124\pm 70$ & $   4.217\pm 0.10$ & $ 1.41^{+0.24}_{-0.16}$ & $  10.151\pm    0.220$ & $   7.206\pm     0.87$ & $    6.60\pm     1.01$ & $0.71$ \\
   318 &    8156120 &  $ 6338\pm 70$ & $   4.169\pm 0.10$ & $ 1.56^{+0.25}_{-0.19}$ & $   5.117\pm    0.127$ & $  15.665\pm     1.85$ & $   13.50\pm     1.00$ & $0.69$ \\
   323 &    9139084 &  $ 5528\pm 70$ & $   4.720\pm 0.10$ & $ 0.89^{+0.04}_{-0.03}$ & $   7.621\pm    0.023$ & $   5.905\pm     0.22$ & $    4.70\pm     1.03$ & $0.80$ \\
   333 &   10337258 &  $ 6208\pm 70$ & $   4.418\pm 0.10$ & $ 1.19^{+0.12}_{-0.08}$ & $   6.909\pm    0.150$ & $   8.864\pm     0.60$ & $    8.30\pm     1.00$ & $0.63$ \\
   372 &    6471021 &  $ 5815\pm 70$ & $   4.597\pm 0.10$ & $ 0.96^{+0.06}_{-0.04}$ & $  11.887\pm    0.087$ & $   4.115\pm     0.20$ & $    3.00\pm     1.16$ & $0.89$ \\
   590 &    9782691 &  $ 5975\pm 70$ & $   4.415\pm 0.10$ & $ 1.08^{+0.11}_{-0.08}$ & $  13.482\pm    0.380$ & $   4.120\pm     0.27$ & $    3.70\pm     1.08$ & $0.81$ \\
   620 &   11773022 &  $ 5673\pm 70$ & $   4.697\pm 0.10$ & $ 0.92^{+0.05}_{-0.03}$ & $   8.212\pm    0.042$ & $   5.680\pm     0.23$ & $    5.50\pm     1.02$ & $0.69$ \\
   665 &    6685609 &  $ 5969\pm 70$ & $   4.092\pm 0.10$ & $ 1.69^{+0.28}_{-0.26}$ & $  15.703\pm    0.960$ & $   5.451\pm     0.65$ & $    5.30\pm     1.02$ & $0.72$ \\
   673 &    7124613 &  $ 6380\pm 70$ & $   4.466\pm 0.10$ & $ 1.22^{+0.10}_{-0.06}$ & $   4.842\pm    0.200$ & $  12.889\pm     0.59$ & $   12.70\pm     1.00$ & $0.51$ \\
   720 &    9963524 &  $ 5260\pm 70$ & $   4.677\pm 0.10$ & $ 0.82^{+0.04}_{-0.03}$ & $   9.529\pm    0.099$ & $   4.399\pm     0.15$ & $    4.50\pm     1.04$ & $0.73$ \\
   723 &   10002866 &  $ 5314\pm 70$ & $   4.555\pm 0.10$ & $ 0.91^{+0.07}_{-0.04}$ & $  11.060\pm    0.047$ & $   4.216\pm     0.24$ & $    4.20\pm     1.05$ & $0.75$ \\
   896 &    7825899 &  $ 4973\pm 70$ & $   3.945\pm 0.10$ & $ 2.09^{+0.31}_{-0.28}$ & $  25.132\pm    0.202$ & $   4.215\pm     0.58$ & $    1.10\pm     1.92$ & $0.97$ \\
   975 &    3632418 &  $ 6202\pm 70$ & $   4.079\pm 0.10$ & $ 1.58^{+0.22}_{-0.19}$ & $  12.553\pm    0.170$ & $   6.388\pm     0.77$ & $    7.30\pm     1.01$ & $0.58$ \\
  1353 &    7303287 &  $ 5989\pm 70$ & $   4.596\pm 0.10$ & $ 1.03^{+0.07}_{-0.05}$ & $   8.789\pm    0.059$ & $   5.960\pm     0.29$ & $    5.80\pm     1.01$ & $0.68$ \\
  1445 &   11336883 &  $ 6318\pm 70$ & $   4.255\pm 0.10$ & $ 1.38^{+0.20}_{-0.15}$ & $   5.389\pm    0.102$ & $  13.209\pm     1.48$ & $   13.80\pm     1.00$ & $0.53$ \\
  1612 &   10963065 &  $ 6089\pm 70$ & $   4.305\pm 0.10$ & $ 1.17^{+0.14}_{-0.11}$ & $  12.650\pm    0.300$ & $   4.719\pm     0.42$ & $    2.80\pm     1.21$ & $0.93$ \\
  1616 &    9015738 &  $ 6042\pm 70$ & $   4.265\pm 0.10$ & $ 1.30^{+0.19}_{-0.14}$ & $  11.711\pm    1.550$ & $   5.763\pm     0.61$ & $    5.30\pm     1.02$ & $0.74$ \\
  1621 &    5561278 &  $ 6079\pm 70$ & $   4.038\pm 0.10$ & $ 1.68^{+0.26}_{-0.22}$ & $  17.975\pm    0.420$ & $   4.758\pm     0.57$ & $    5.90\pm     1.01$ & $0.61$ \\
  1628 &    6975129 &  $ 6223\pm 70$ & $   4.484\pm 0.10$ & $ 1.17^{+0.10}_{-0.06}$ & $   5.731\pm    0.650$ & $  10.561\pm     0.86$ & $   10.40\pm     1.00$ & $0.57$ \\
  1800 &   11017901 &  $ 5620\pm 70$ & $   4.686\pm 0.10$ & $ 0.91^{+0.05}_{-0.03}$ & $   6.536\pm    0.046$ & $   7.068\pm     0.27$ & $    6.90\pm     1.01$ & $0.63$ \\
  1825 &    5375194 &  $ 5390\pm 70$ & $   4.687\pm 0.10$ & $ 0.86^{+0.04}_{-0.03}$ & $  10.337\pm    0.194$ & $   4.231\pm     0.13$ & $    4.50\pm     1.04$ & $0.72$ \\
  1839 &    5856571 &  $ 5517\pm 70$ & $   4.665\pm 0.10$ & $ 0.90^{+0.05}_{-0.03}$ & $   6.280\pm    0.020$ & $   7.309\pm     0.33$ & $    7.10\pm     1.01$ & $0.63$ \\
  1883 &   11758544 &  $ 6059\pm 70$ & $   4.148\pm 0.10$ & $ 1.49^{+0.23}_{-0.19}$ & $  11.632\pm    0.280$ & $   6.495\pm     0.76$ & $    6.60\pm     1.01$ & $0.66$ \\
  1886 &    9549648 &  $ 6200\pm 70$ & $   4.195\pm 0.10$ & $ 1.44^{+0.22}_{-0.18}$ & $   7.641\pm    0.499$ & $   9.354\pm     0.94$ & $   10.50\pm     1.00$ & $0.50$ \\
  1958 &    9836149 &  $ 5785\pm 70$ & $   4.575\pm 0.10$ & $ 0.97^{+0.07}_{-0.05}$ & $  10.501\pm    0.111$ & $   4.741\pm     0.25$ & $    4.10\pm     1.05$ & $0.80$ \\
  2002 &   10024701 &  $ 6004\pm 70$ & $   4.499\pm 0.10$ & $ 1.06^{+0.09}_{-0.06}$ & $  10.083\pm    0.610$ & $   5.400\pm     0.31$ & $    4.80\pm     1.03$ & $0.76$ \\
  2026 &   11923284 &  $ 5994\pm 70$ & $   4.516\pm 0.10$ & $ 1.01^{+0.08}_{-0.06}$ & $   9.775\pm    0.120$ & $   5.320\pm     0.31$ & $    4.50\pm     1.04$ & $0.79$ \\
  2035 &    9790806 &  $ 5557\pm 70$ & $   4.670\pm 0.10$ & $ 0.90^{+0.05}_{-0.03}$ & $   7.113\pm    0.053$ & $   6.453\pm     0.28$ & $    6.00\pm     1.01$ & $0.69$ \\
  2109 &   11499228 &  $ 6084\pm 70$ & $   4.093\pm 0.10$ & $ 1.59^{+0.26}_{-0.20}$ & $  11.197\pm    0.090$ & $   7.324\pm     1.04$ & $    7.40\pm     1.01$ & $0.66$ \\
  2110 &   11460462 &  $ 6385\pm 70$ & $   4.260\pm 0.10$ & $ 1.40^{+0.20}_{-0.13}$ & $   4.326\pm    0.066$ & $  16.689\pm     1.80$ & $   16.40\pm     1.00$ & $0.56$ \\
  2111 &    8612275 &  $ 5604\pm 70$ & $   4.589\pm 0.10$ & $ 0.93^{+0.06}_{-0.04}$ & $  10.233\pm    0.054$ & $   4.656\pm     0.25$ & $    3.70\pm     1.08$ & $0.84$ \\
  2273 &    9717943 &  $ 6028\pm 70$ & $   4.271\pm 0.10$ & $ 1.29^{+0.20}_{-0.13}$ & $  12.507\pm    0.240$ & $   5.357\pm     0.59$ & $    4.50\pm     1.04$ & $0.80$ \\
  2403 &    2142522 &  $ 6125\pm 70$ & $   4.323\pm 0.10$ & $ 1.25^{+0.15}_{-0.11}$ & $  10.443\pm    0.130$ & $   6.156\pm     0.61$ & $    5.30\pm     1.02$ & $0.76$ \\
  2545 &    9696358 &  $ 6197\pm 70$ & $   4.035\pm 0.10$ & $ 1.83^{+0.30}_{-0.26}$ & $  17.069\pm    0.470$ & $   5.454\pm     0.71$ & $    6.80\pm     1.01$ & $0.58$ \\
  2555 &    5350244 &  $ 6144\pm 70$ & $   4.349\pm 0.10$ & $ 1.18^{+0.14}_{-0.10}$ & $   8.900\pm    0.110$ & $   6.844\pm     0.64$ & $    6.20\pm     1.01$ & $0.71$ \\
  2593 &    8212002 &  $ 6212\pm 70$ & $   4.259\pm 0.10$ & $ 1.35^{+0.21}_{-0.14}$ & $   8.690\pm    0.180$ & $   8.061\pm     0.93$ & $    7.40\pm     1.01$ & $0.69$ \\
  2632 &   11337566 &  $ 6237\pm 70$ & $   4.059\pm 0.10$ & $ 1.68^{+0.30}_{-0.23}$ & $  11.054\pm    0.210$ & $   7.856\pm     1.09$ & $   10.30\pm     1.00$ & $0.47$ \\
  2675 &    5794570 &  $ 5755\pm 70$ & $   4.632\pm 0.10$ & $ 0.96^{+0.06}_{-0.04}$ & $   6.071\pm    0.061$ & $   8.094\pm     0.37$ & $    7.70\pm     1.00$ & $0.63$ \\
  2678 &    6779260 &  $ 5415\pm 70$ & $   4.703\pm 0.10$ & $ 0.86^{+0.05}_{-0.03}$ & $   6.193\pm    0.017$ & $   7.099\pm     0.30$ & $    6.80\pm     1.01$ & $0.65$ \\
  2722 &    7673192 &  $ 6119\pm 70$ & $   4.381\pm 0.10$ & $ 1.14^{+0.13}_{-0.09}$ & $   9.069\pm    0.210$ & $   6.480\pm     0.48$ & $    6.30\pm     1.01$ & $0.67$ \\
  2961 &   10471515 &  $ 6069\pm 70$ & $   4.311\pm 0.10$ & $ 1.25^{+0.17}_{-0.12}$ & $  11.898\pm    0.440$ & $   5.394\pm     0.48$ & $    5.20\pm     1.02$ & $0.72$ \\
  3060 &   11019987 &  $ 6189\pm 70$ & $   4.055\pm 0.10$ & $ 1.61^{+0.22}_{-0.19}$ & $  13.174\pm    0.350$ & $   6.223\pm     0.67$ & $    6.20\pm     1.01$ & $0.67$ \\
  3681 &    2581316 &  $ 6194\pm 70$ & $   4.292\pm 0.10$ & $ 1.29^{+0.19}_{-0.12}$ & $   7.789\pm    0.475$ & $   8.455\pm     0.70$ & $    8.10\pm     1.00$ & $0.63$ \\
  3683 &   10795103 &  $ 6338\pm 70$ & $   4.274\pm 0.10$ & $ 1.32^{+0.17}_{-0.12}$ & $   6.899\pm    1.450$ & $  10.039\pm     1.49$ & $   10.70\pm     1.00$ & $0.58$ \\
  3835 &    2581554 &  $ 5013\pm 70$ & $   4.703\pm 0.10$ & $ 0.78^{+0.03}_{-0.02}$ & $   7.877\pm    0.029$ & $   5.006\pm     0.16$ & $    4.90\pm     1.03$ & $0.71$ \\
  4160 &    7610663 &  $ 5766\pm 70$ & $   4.497\pm 0.10$ & $ 1.03^{+0.10}_{-0.06}$ & $  10.418\pm    0.040$ & $   5.101\pm     0.37$ & $    3.50\pm     1.10$ & $0.88$ \\
  4246 &    5177859 &  $ 5723\pm 70$ & $   4.624\pm 0.10$ & $ 0.99^{+0.06}_{-0.05}$ & $   7.227\pm    0.057$ & $   7.018\pm     0.32$ & $    6.40\pm     1.01$ & $0.69$ \\
  4276 &    6026924 &  $ 6300\pm 70$ & $   3.982\pm 0.10$ & $ 1.99^{+0.31}_{-0.28}$ & $  16.212\pm    0.271$ & $   6.222\pm     0.87$ & $    8.20\pm     1.00$ & $0.52$ \\
  4411 &    5281113 &  $ 6043\pm 70$ & $   4.348\pm 0.10$ & $ 1.12^{+0.13}_{-0.10}$ & $  11.754\pm    1.390$ & $   4.896\pm     0.43$ & $    4.20\pm     1.05$ & $0.80$ \\
\enddata
\end{deluxetable*}

\begin{deluxetable*}{ccccccccc}
\tabletypesize{\small}
\tablecaption{Stars previously identified as possibly
  misaligned\label{tbl:prev-cands}}
\tablewidth{0pt}

\tablehead{
  \colhead{KOI} &
  \colhead{KIC} &
  \colhead{$T_{\rm eff}$} &
  \colhead{$\log g$} &
  \colhead{Radius} &
  \colhead{Rotation period\tablenotemark{a}} &
  \colhead{$v$} &
  \colhead{$v\sin i$} &
  \colhead{$N_\sigma$\tablenotemark{b}} \\
  \colhead{number} &
  \colhead{number} &
  \colhead{[K]} &
  \colhead{[cgs]} &
  \colhead{[$R_\star$]} &
  \colhead{[days]} &
  \colhead{[km~s$^{-1}$]} &
  \colhead{[km~s$^{-1}$]} &
  \colhead{}
}
\renewcommand{\arraystretch}{1.0}
\startdata
   261 &    5383248 &  $ 5750\pm 70$ & $   4.504\pm 0.10$ & $ 0.99^{+0.08}_{-0.05}$ & $   15.01\pm     0.10$ & $    3.34\pm     0.21$ & $    0.20\pm     2.00$ & $   1.57$ \\
   323 &    9139084 &  $ 5528\pm 70$ & $   4.720\pm 0.10$ & $ 0.89^{+0.04}_{-0.03}$ & $    7.62\pm     0.00$ & $    5.87\pm     0.24$ & $    4.70\pm     1.03$ & $   1.11$ \\
   377 &    3323887 &  $ 5787\pm 70$ & $   4.473\pm 0.10$ & $ 1.02^{+0.10}_{-0.06}$ & $   16.76\pm     0.09$ & $    3.07\pm     0.23$ & $    1.10\pm     1.92$ & $   1.02$ \\
   988 &    2302548 &  $ 5121\pm 70$ & $   4.687\pm 0.10$ & $ 0.81^{+0.04}_{-0.03}$ & $   12.43\pm     0.01$ & $    3.28\pm     0.13$ & $    3.30\pm     1.12$ & $  -0.02$ \\
  1890 &    7449136 &  $ 6119\pm 70$ & $   4.190\pm 0.10$ & $ 1.43^{+0.24}_{-0.18}$ & \nodata & \nodata & $    7.20\pm     1.01$ & \nodata \\
  2002 &   10024701 &  $ 6004\pm 70$ & $   4.499\pm 0.10$ & $ 1.06^{+0.09}_{-0.06}$ & $   10.08\pm     0.05$ & $    5.32\pm     0.36$ & $    4.80\pm     1.03$ & $   0.49$ \\
  2026 &   11923284 &  $ 5994\pm 70$ & $   4.516\pm 0.10$ & $ 1.01^{+0.08}_{-0.06}$ & $    9.77\pm     0.05$ & $    5.25\pm     0.32$ & $    4.50\pm     1.04$ & $   0.70$ \\
  2261 &    3734418 &  $ 5176\pm 70$ & $   4.696\pm 0.10$ & $ 0.81^{+0.04}_{-0.03}$ & $   11.43\pm     0.00$ & $    3.58\pm     0.14$ & $    3.10\pm     1.15$ & $   0.42$
\enddata

\tablenotetext{a}{A blank entry indicates that the photometric
  rotation period was not deemed to be reliable, i.e., the catalogs of
  \citet{Mazeh+2015} and \citet{Angus+2017} do not report consistent
  results.}
  
\tablenotetext{b}{Defined as $v-v\sin i$ divided by the quadrature sum
  of the uncertainties in $v$ and $v\sin i$.}

\end{deluxetable*}

\end{document}